\documentclass[showpacs,lettersize,twocolumn]{revtex4}
\usepackage{epsfig}
\usepackage{amsmath}
\usepackage{graphicx}

\begin{document}

\title{Experimental Observation of Dramatic Differences in the\\
Dynamic Response of Newtonian and Maxwellian Fluids}
\author{J. R. Castrej\'on-Pita}
\author{J. A. del R{\'\i}o$^{\dag}$}
\author{A. A. Castrej\'on-Pita}
\author{G. Huelsz}

\address{Centro de Investigaci\'on en Energ{\'\i}a, UNAM\\
Apdo. Postal 34, 62580 Temixco, Mor. M\'exico.\\
$^{\dag}$ {\it email}: antonio@servidor.unam.mx}

\begin{abstract}
An experimental study of the dynamic response of a Newtonian fluid
and a Maxwellian fluid under an oscillating pressure gradient is
presented. Laser Doppler anemometry is used in order to determine
the velocity of the fluid inside a cylindrical tube. In the case
of the Newtonian fluid, the dissipative nature is observed. In the
dynamic response of the Maxwellian fluid an enhancement at the
frequencies predicted by theory is observed. \vskip1cm
\end{abstract}

\maketitle

\section{Introduction}

Since the concept of dynamic permeability was proposed several
years ago \cite{johnson,Sheng}, it has been extensively used in
the study of practical problems such as petroleum
recovery~\cite{spe}, soil-ground water transport \cite{Salem},
fluids flowing in porous media \cite{dongen,chapman}, acoustic
waves in porous media~\cite{geo}, wave propagation in
foams~\cite{apl}, fluid circulation in biological
systems~\cite{thurs}, etc. The use of the dynamic permeability as
originally proposed~\cite{johnson} however, is not quite correct
for some of these systems since the fluid does not behave as a
Newtonian fluid but shows viscoelastic features. Theoretical
analyzes of viscoelastic fluids have recently predicted an
interesting enhancement in the dynamic permeability of several
orders of magnitude in comparison with the static
permeability~\cite{Meeting,tipm,pre,Beresnev}. This behavior is
due to the coupling between the elastic behavior of the fluid and
the geometry of the container and is completely different from the
pure dissipative behavior of Newtonian fluids. An increase in the
flow rate of a viscoelastic fluid flowing in a tube under
oscillating conditions was discovered several years
ago~\cite{thurston52,barnes,manero,davies} but up to now, the
experimental measurements in terms of a frequency-dependent
response had not been performed. Some theoretical studies have
explored interesting consequences of the enhancement of the
dynamic permeability in different systems including viscoelastic
fluids~\cite{pre,Beresnev,cuevas,Eugenia,rmf,david,david3,david2}.
The variety of systems where an oscillating pressure gradient can
be imposed to a viscoelastic fluid is wide, for instance, from the
oil recovery problem~\cite{patentoil} to the dynamic analysis of
fluid transport in animals (including the human
body~\cite{blood}). All previous theoretical derivations of the
dynamic permeability for viscoelastic fluids have been made in
terms of the averaged velocity. In this work however, we perform a
detailed analysis by measuring the dynamic response at the center
of a fluid column that is moved in an oscillatory way and its
local velocity \cite{fut} is measured using a laser Doppler
anemometer (LDA). This allows us to test a local, simple model for
both Newtonian and Maxwellian fluids and our results corroborate
that the simple linear model can capture the differences between
the dissipative (Newtonian fluid) and resonant (Maxwellian fluid)
behaviors.

\section{Linear theory}

In the following, we present the pertinent equations used to obtain a local
expression for the Maxwellian fluid response based on the results presented
elsewhere~\cite{pre}. We begin with the \textit{linearized} momentum
equation
\begin{equation}
\rho \ \frac{\partial \mathbf{v}}{\partial t}\ =\ -\ \nabla p\ +\ \nabla
\cdot \overleftrightarrow{\tau },  \label{momentum}
\end{equation}%
and the continuity equation
\begin{equation}
\nabla \cdot \mathbf{v}\ =\ 0,
\end{equation}%
for an incompressible fluid. In the above equations, $\rho $
denotes the mass density of the fluid, ${\mathbf{v}}$ the fluid
velocity, $p$ the pressure, and $\overleftrightarrow{\tau }$
represents the viscous stress tensor. To ensure the validity of
Eq.~(\ref{momentum}) it is necessary to consider fluids with low
Reynolds numbers, in our experimental case: $\ Re\ <\
7$x$10^{-4}.\ $ The constitutive equation of the fluid that we use
is the \textit{linearized} Maxwell model:
\begin{equation}
t_{m}\ \frac{\partial \overleftrightarrow{\tau }}{\partial t}\ =\ -\ \eta \
\nabla \mathbf{v}\ -\ \overleftrightarrow{\tau },
\end{equation}%
where $\eta $ denotes the dynamic viscosity and $t_{m}$ the relaxation time.
It is necessary to stress that the linearized Maxwell model constitutes a
simplification required to obtain simple analytic results that implies
neglecting non-linear terms which may be important under certain
circumstances.

Manipulating the three previous equations, applying the Fourier transform,
and using cylindrical coordinates, we obtain
\begin{equation}
V(r,\omega ) \ = \ - \ \frac{\left( 1-\mathit{i}\omega t_m\right) }{\eta
\beta ^2} \ \left( 1 - \frac{J_0(\beta r)}{J_0(\beta a)} \right) \ \frac{dP}{%
dz},  \label{vel}
\end{equation}
where $\beta = \sqrt{ \rho \left[ \left(t_m \omega \right) ^2 +
\mathit{i} \omega t_m \right]/(\eta t_m )} \ , \ a$ is the
cylinder radius, $P$ and $V$ are the pressure and the velocity in
the frequency domain, respectively, $\omega $ is the angular
frequency, and $J_0$ is the cylindrical Bessel function of zeroth
order. A more detailed derivation of these expressions can be
found in~\cite{pre,rmf}.

Previous theoretical analyzes of the dynamic permeability of
viscoelastic fluids deal with the above equations using an average
velocity for the fluid. In what follows, we shall analyze the flow
from a different point of view, one which is based on a ``local
response'' defined as the response of the fluid to a pressure
gradient in a particular position inside the tube. This is
required by the LDA that essentially measures quasi-local particle
velocities. Thus, in order to analyze the local dynamic response,
and to facilitate the measurements, we evaluate Eq. (\ref{vel}) at
the center of the cylinder ($r=0$ in cylindrical coordinates)
where the amplitude of the velocity is maximum:

\begin{equation} V(\omega
)\ =\ -\ \frac{\left( 1-\mathit{i}\omega t_{m}\right) }{\eta \beta
^{2}}\ \left( 1-\frac{1}{J_{0}(\beta a)}\right) \ \frac{dP}{dz}.
\label{velo}
\end{equation}%
The local dynamic response is thus defined as:
\begin{equation}
\xi \ =\ -\ \eta \ \frac{V}{dP/dz},  \label{k}
\end{equation}%
where we have followed Darcy's generalized
equation~\cite{Meeting}. Substituting the velocity given by Eq.
(\ref{velo}), we obtain:
\begin{equation}
\xi (\omega )\ =\ \frac{\left( 1-\mathit{i}\omega t_{m}\right) }{\beta ^{2}}%
\left( 1-\frac{1}{J_{0}(\beta a)}\right) .  \label{kteo}
\end{equation}%
From this expression one can recover the Newtonian behavior with
the substitution: $t_{m}=0.$ In order to compare the last
expression with experimental results, we need an expression for
the pressure gradient. In our experimental case, we have a
harmonic oscillating column of fluid being moved by a piston
coupled to a motor with adjustable frequency. Thus, the pressure
gradient can be easily described by:
\begin{equation}
\frac{dp(t)}{dz}\ =\  \rho \ z_{0}\ \omega ^{2}\ \sin (\omega t),
\label{8}
\end{equation}%
which represents an oscillatory movement in the $z$ direction with
a displacement amplitude equal to $z_{0}$. We shall also measure
the root mean square velocity, $v_{rms}$, since the range of
frequencies that can be resolved in this case is much wider than
the range given by the Fourier transform of the velocity data. The
experimental dynamic response of a fluid, ($\xi ^{\exp }(\omega
)$), is thus given by:
\begin{equation}
\xi ^{\exp }\ =\ \eta \ \frac{v_{rms}}{dp_{rms}/dz}\ =\
\frac{\sqrt{2}\ \eta \ v_{rms}}{\rho \ z_{0}\ \omega ^{2}}.
\label{10}
\end{equation}

Now, since the experimental velocity is harmonic, we can directly
compare the value of the local dynamic response obtained using the
previous expression for each experiment with the theoretical
prediction from Eq. (\ref{kteo}).

\section{Experimental System}

The experimental device shown in Fig. 1, consists of a vertical
cylinder filled with a fluid. The oscillating movement required to
have a harmonic oscillating pressure gradient, \textit{c. f.}
Eq.~(\ref{8}), was produced with a piston that closes the base of
the cylinder and is driven by a motor of variable frequency; as
previously mentioned, the local velocity of the fluid is measured
with a laser Doppler anemometer.

\begin{figure}[tbp]
\begin{center}
\epsfig{width=6cm,file=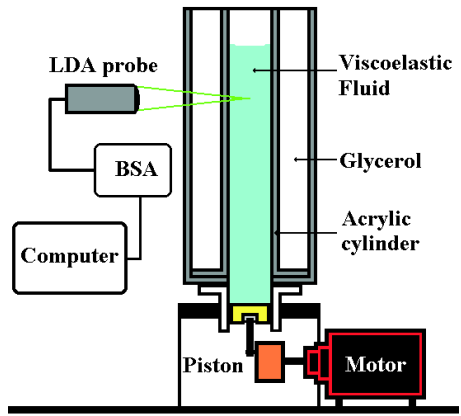}
\end{center}
\caption{Schematic view of the setup used for the study of the
dynamic response with an LDA measuring system} \label{fig:1}
\end{figure}

The cylindrical tube, made of transparent acrylic, has an inner
diameter of $5 \ cm$ and a length of $50 \ cm.$ In its lower end,
this tube is joined by a clamp to a stainless steel piston skirt
of the same inner diameter. Within the stainless steel skirt,
there is a Teflon piston that is moved by a \textit{Siemens} motor
of $1 \ Hp$, regulated by a \textit{Varispeed 606PC3} system. This
allows for a frequency of oscillation that can be varied from
$1.5$ to $200 \ Hz.$ The acrylic cylinder over the skirt is
contained in a square recipient with parallel walls of transparent
acrylic and filled with glycerol having a refractive index similar
to the one of the acrylic; this is in order to avoid the
cylindrical aberration due to the fact that the cylinder acts as a
lens. In this and other cases, when there are cylindrical
configurations, these aberrations could cause a null intersection
between the LDA beams or could result in the backscattering signal
not being received by the photomultiplier. The laser Doppler
anemometer (LDA) used for the local velocity measurements is a
widely known system~\cite{Adrian}. Our particular LDA consists of
a probe (\textit{Dantec FiberFlow 60x20}), a photomultiplier, a
Burst Spectrum Analyzer (\textit{Dantec BSA 57N11}), and an argon
laser (\textit{Spectra Physics 177}) emitting in four different
wavelengths; only the $488 \ nm \ $ line ($200 \ mW$) is used, and
the data are processed in a PC with the \textit{Dantec
FlowManager} software. The measurement volume is of $0.64 \ \times
\ 0.075 \ \times \ 0.075 \ mm \ $ located at the center of the
acrylic cylinder. The LDA probe was set into a positioning system
(precision: $1 \ \mu m$) placed on a different table from the rest
of the setup to prevent oscillations that could alter the
experimental data. The \textit{rms}-velocity is calculated from
the collected data.

Commercial glycerol was used as the Newtonian fluid to be tested;
it has a viscosity of $\ 1 \ Pa \cdot$s and a density of $\ 1250 \
kg/m^3. \ $ The well-known cetylpyridinium chloride and sodium
salicylate solution (CPyCl/NaSal,
$60$/$100$)~\cite{Micelas,Hoffman} was used as the Maxwellian
counterpart; its properties are $\rho = 1050 \ kg/m^3, \ \eta = 60
\ Pa \cdot $s, and $\ t_m = 1.9 \ s$~\cite{Lourdes}. All the
measurements were made within the $\ 25 \pm 0.5^o \ C \ $ interval
in order to keep constant the properties of CPyCl/NaSal solution
\cite{Lourdes}. The properties of this solution and the diameter
of the tube give a Deborah's number ($\alpha =\rho a^{2}$ / $\eta
t_{m}$) of $\alpha =0.0058$. This value is much smaller than the
critical value $\alpha <11.64 $ for the appearance of resonances
predicted by the theoretical model \cite{pre,rmf}. The
experimental frequency range is kept within the low Reynolds
number regime by setting the piston movement amplitude to $\ 0.8
\pm 0.05 \ mm$ (which assures that $\ Re < 7 \times 10^{-4}$).
Under these circumstances we checked that the system can be described
by a linearized balance momentum equation.
\textit{Dantec} $\ 20 \ \mu m \ $ polyamid spheres were used as
seeding particles; they remain suspended for a long time and cause
a minimal standard deviation in the velocity data since the size
of the particles is uniform. We can assure that the particles
follow the flow because, using Stokes's law \cite{Drain}, the
obtained limit frequency
\begin{equation}
f<0.1\frac{\eta }{\rho _{P}R^{2}},
\end{equation}
 for which the particles follow an oscillation
 with a deviation up to 1 \%, in amplitude, is much greater than the experimental
frequencies (between 1.5 Hz and 15 Hz);  $f = 9.7$x$10^{7}$ Hz for
glycerol and $f = 5.8$x$10^{9}$ Hz for the CPyCl/NaSal solution.
The particle density, $\rho _{P}$, is 10.3 kg/m$^{3}$ and  the
radius of a particle, $R$, is 1x10$^{-5}$ m \cite{Dantec}.

To avoid transient perturbations, measurements were taken
approximately $\ 5 \ minutes$ after every change in the frequency.

\section{Results}

For each frequency $\ 2,000 \ $ velocity data were acquired with
the LDA at the cylinder axis and $\ 40 \ cm \ $ above the piston,
where edge effects are negligible. Experiments were carried out in
the oscillating frequency range $[1.5, 15] \pm 0.1 \ Hz $. 
In this frequency range the viscoelastic shear modulus is
nearly real-valued (lossy elastic solid). The
calculated \textit{rms}-velocity from the LDA measurements for
CPyCl/NaSal and for glycerol are shown in Fig 2, where one
observes that the measured Maxwellian velocity is dramatically
different from the Newtonian velocity in the studied range.
Clearly the \textit{rms}-velocity increases with the frequency in
both cases, but the velocity of the viscoelastic fluid shows
well-defined peaks. An important point to be stressed is the fact
that the velocity in the glycerol case is a linear function with
respect to the frequency which means a good quality of the
mechanical piston movement. Moreover, this confirms that the
movement of the piston is transmitted directly to the fluid.

\begin{figure}[tbp]
\begin{center}
\epsfig{width=9cm,file=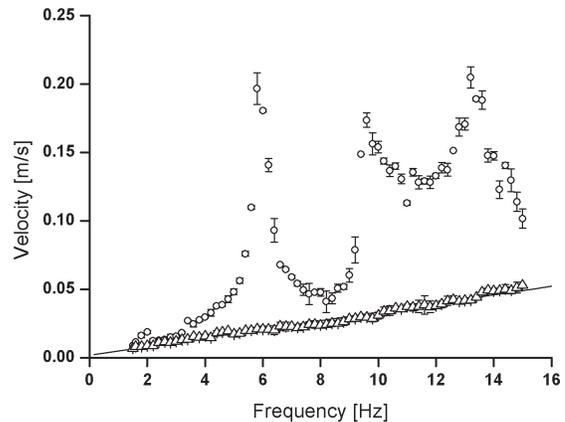}
\end{center}
\caption{Local \textit{rms}-velocity at the center of the
cylinder. The circles correspond to the viscoelastic fluid and the
triangles to the Newtonian fluid} \label{fig:2}
\end{figure}

Originally inferred by the pioneers in the analysis of the dynamic
permeability~\cite{johnson,Sheng}, the dissipative nature of the
glycerol response is clearly seen in Fig. 3, where the glycerol
dynamic response and the frequency have been scaled, the first one
by the value of the dynamic response at $\ \omega = 0 \ $  and the
second by the viscous time $\ \tau = a^2\rho/\eta. \ $ Also shown
in Fig. 3 are the theoretical predictions derived from Eq. (7)
(magnitude of $\xi$ with $t_{m}=0$).

\begin{figure}[tbp]
\begin{center}
\epsfig{width=9cm,file=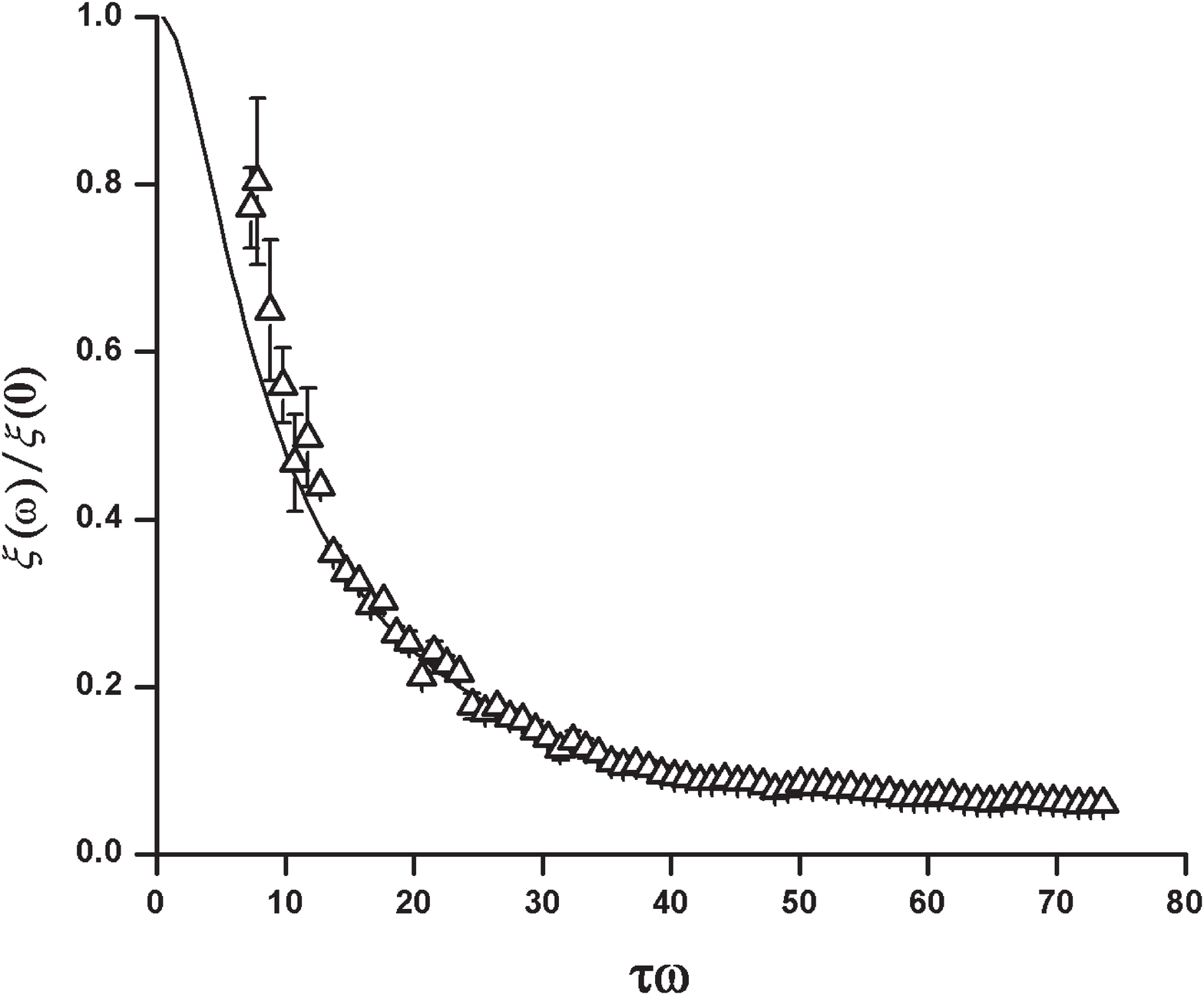}
\end{center}
\caption{ Dynamic response for glycerol. Experimental values are
shown by triangles (Eq. 9) and the line represents the theoretical
prediction (magnitude of $\xi$ from Eq. 7 with $t_{m}=0$)}.
\label{fig:3}
\end{figure}

Fig. 4 contains the dynamic response of the CPyCl/NaSal solution.
Experimental values are plotted together with theoretical
predictions (Eq.~\ref{kteo}). A dramatic change in the behavior of
the dimensionless Maxwellian response with respect to the
dissipative Newtonian case is observed, namely resonant
frequencies appear. The agreement with Eq.~(\ref{kteo}) is
manifest: the positions of the peaks in Eq.~(\ref{kteo}) and in
the experimental results occur at approximately the same values of
the frequency $\omega $, even though the amplitudes differ. We
need to recall that the theory used for the prediction is a linear
approximation and neglects the convective term and the possible
non linearities in the viscoelastic behavior of the fluid.
Although we have selected the viscoelastic fluid in order to
maintain the physical conditions where resonance behavior appears,
we were unable to assure the linear characteristic of the
viscoelastic fluid. According to \cite{Lourdes} the CPCl/NaSal
solution behaves as a Maxwellian fluid if the shear rate is
$\gamma <0.6$ s$^{-1}$. Transforming Eq. 4 to the time space and
taking the radial derivative we have calculated the shear rate
that depends on the radial coordinate and time. The maximum local
shear rate increases with frequency, \textit{e. g.}
$\gamma_{2Hz}<1.5$ s$^{-1},$ $\gamma _{6.5Hz}<21$ s$^{-1},$
$\gamma_{10Hz}<70$ s$^{-1}$, where the subscripts indicate the
frequency. Then, one of the possible causes of the disagreement in
the response amplitude is because the fluid was locally under
higher shear rates than its Maxwellian limit. This is in agreement
with the fact that if we increase the frequency, then we find
larger differences between theory and experiment. Another reason
for theoretical and experimental differences could be compressible
effects in the viscoelastic flow that also increase with
frequency.

 These results constitute the first experimental
evidence of the resonant behavior of a viscoelastic fluid and
shows the validity of the theory ~\cite{pre,rmf}.

\begin{figure}[tbp]
\begin{center}
\epsfig{width=9cm,file=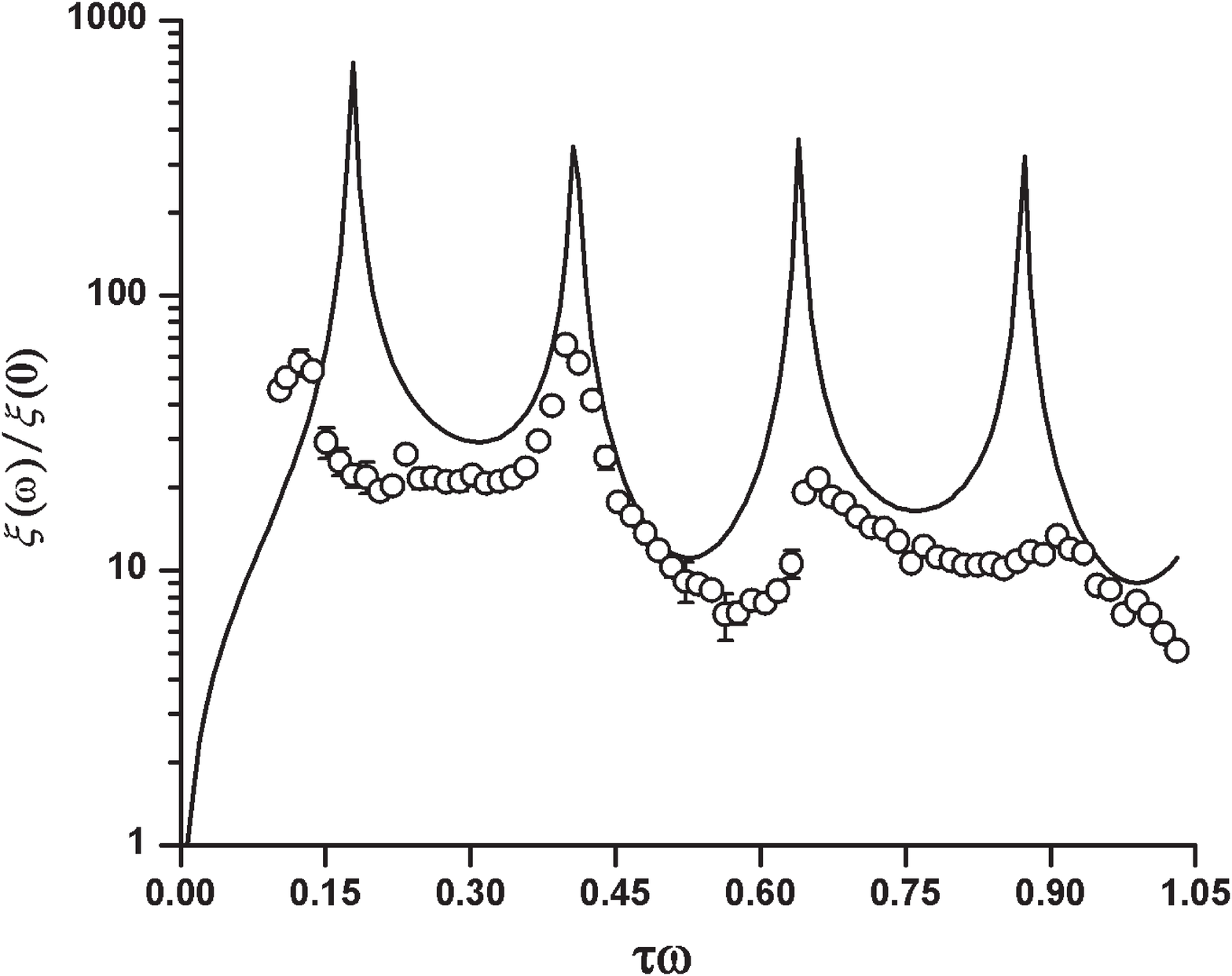}
\end{center}
\caption{Dynamic response for the CPyCl/NaSal solution.
Experimental values are shown by circles (Eq. 9) and the line
represents the theoretical prediction (magnitude of $\xi$ from Eq.
7)}. \label{fig:4}
\end{figure}

\section{Conclusions}

The local dynamic response of a fluid (either Newtonian or
Maxwellian) has been developed following previous work
\cite{johnson,Sheng,tipm,pre}. This theoretical expression is
compared with experimental LDA measurements in a
harmonically-oscillating pressure system. The expected dissipative
behavior of Newtonian fluids is confirmed. On the other hand and
although the theoretical approximation presented is a drastic
simplification of the real viscoelastic behavior, it is still
capable of reproducing the resonant behavior of the dynamic
response of a Maxwellian fluid. The resonances appearing in the
viscoelastic response, as expected, are significantly higher than
the monotonic decay in the Newtonian response. In both cases, the
qualitatively agreement between theoretical predictions and the
experimental results presented here is evident; the quantitative
differences are probably due to non-linear effects of the
viscoelastic fluid used. Our experimental results show that the
dynamic response of viscoelastic fluids clearly exceeds the static
response for some specific frequencies, in fair agreement with the
theory for such systems. Finally, we want to stress that the
results obtained in this work have a wide spectrum of
applications, from oil recovery problems~\cite{patentoil} to the
dynamical analysis of biological fluids~\cite{blood,lung,lung2}.

\section{Acknowledgments}

This work has been partially supported by CONACyT (32707-U,
38538). JRCP acknowledges financial support from DGAPA-UNAM
(IN101100). The authors acknowledge the invaluable help of J.
P\'erez, L. de Vargas and A. M\'endez in characterizing the physical
properties of the CPyCl/NaSal solution; M. L\'opez de Haro for the
encouragement to perform these experiments,
E. Corvera and A. Sarmiento for helpful
discussions.



\begin{thebibliography}{99}
\bibitem{fut} The word local as used herewith means being quasi-local since
the LDA measures particle velocities in a very small, very small
volume.
\end{thebibliography}
\end{document}